\documentstyle[11pt,paspconf,epsfig]{article}

\begin{document}

\title{Galaxy Interactions}

\author{Joshua E. Barnes}

\affil{Institute for Astronomy, University of Hawai`i,
       2680 Woodlawn Drive, Honolulu, Hawai`i, 96822, USA}

\begin{abstract}
Interacting galaxies are a natural arena for studies of {\it
non\/}-equilibrium stellar dynamics, gas dynamics, and thermodynamics.
Only galaxy formation itself is as deeply concerned with as many
different aspects of dynamics, and the connection between interactions
and the formation of galaxies is probably no coincidence.  This review
discusses tidal interactions, halos and orbit decay, dissipative
effects in galaxy interactions, properties of merger remnants, and
origins of starbursts.
\end{abstract}

\section{Simulating Tidal Interactions}

Why simulate interacting galaxies?  First, simulations can {\it test
theoretical ideas\/}.  Second, detailed simulations may help in
gaining {\it insight into real systems\/}.  Third, simulations may
{\it constrain galaxy parameters\/} such as dark halo masses.

Simulation is not a straightforward business.  A dynamical model
specifies the distribution function $f({\bf r}, {\bf v})$, which
depends on {\it six\/} variables.  Observations, at best, yield $f(X,
Y, V_{\rm Z})$, a function of just three variables: two coordinates on
the plane of the sky, and a line of sight velocity.  Thus simulations
are underdetermined; further constraints are needed to make progress.
In cosmology, one may stipulate that the observed structures grew from
a linear density field $\delta \rho({\bf r}) / \rho$ which depends on
three coordinates; this is how the ``least action'' method (Peebles
1994) can yield well-determined results.  But in studying interacting
galaxies we want to understand the {\it stellar\/} distribution, and
the stars did not evolve from linear initial conditions!

So in simulating interacting galaxies, the practice has been to build
equilibrium models and drop them towards each other.  This approach
seems to work provided that the galaxy models and their trajectories
are cosmologically plausible.  One example is NGC~7252, which Hibbard
\& Mihos (1994) simulated successfully as the result of a {\it
direct\/} parabolic encounter of two disk galaxies; an earlier attempt
to reproduce this system with a {\it retrograde\/} encounter (Borne \&
Richstone 1991) required the galaxies to start on implausibly tight
circular orbits and proved inconsistent with subsequent HI
observations.

\subsection{Towards a model of the Antennae}

The Antennae galaxies (NGC~4038/9) are fast becoming the ``Rosetta
stone'' of interacting systems; detailed observations in almost every
waveband from $21 {\rm\, cm}$ to X-rays provide a remarkably complete
picture of the behavior of interstellar material and star formation in
the earlier stages of a galactic merger.  These galaxies have also
long been a favorite of N-body experimenters.  But until recently, the
available line-of-sight velocity data were not good enough to support
detailed simulations.  New VLA observations (Hibbard, van der Hulst,
\& Barnes, in preparation) offer the chance to refine existing models.
Goals for an improved model of the Antennae include:

\begin{enumerate}

\item Matching the observed velocity field.  The radial velocities of
the two galaxies differ by only $\sim 40 {\rm\,km\,sec^{-1}}$.  To
produce this, the galaxies must either be near apocenter, or falling
together almost perpendicular to our line-of-sight.

\item Reconciling the adopted orbit with cosmological expectations.
Simulations by Toomre \& Toomre (1972) and Barnes (1988) adopted
elliptical ($e \simeq 0.6$) orbits; parabolic orbits seem more
plausible.

\item Reproducing the gas-rich ring in NGC~4038.  This ring, clearly
seen in maps of HI as well as in mid-IR (Mirabel et al. 1998),
contains many luminous young star clusters (Whitmore \& Schweizer
1995).

\item Explaining the ``overlap region''.  Recent ISO maps show this
dusty region is brighter than either disk in mid-IR wavebands (eg.,
Mirabel et al. 1998).

\end{enumerate}

Goals one and two involve adjusting the orbit, the viewing angle, and
the orientations of the two disks.  To rapidly explore this vast
parameter space, I run ``semi-consistent'' simulations in which each
galaxy is represented by a self-gravitating spheroid with a number of
embedded test particle disks; the two disks best matching the
observations are selected interactively after the calculation has run.
Starting with orbits as eccentric as $e = 0.8$, this technique yields
models which roughly reproduce the velocity field as well as the
crossed-tail morphology of NGC~4038/9.  But still less than
satisfactory are the shapes of the gently curving tails and the
orientations of their parenting disks; experiments are under way to
study these problems and make models with parabolic initial orbits.

\begin{figure}
\begin{center}
\epsfig{figure=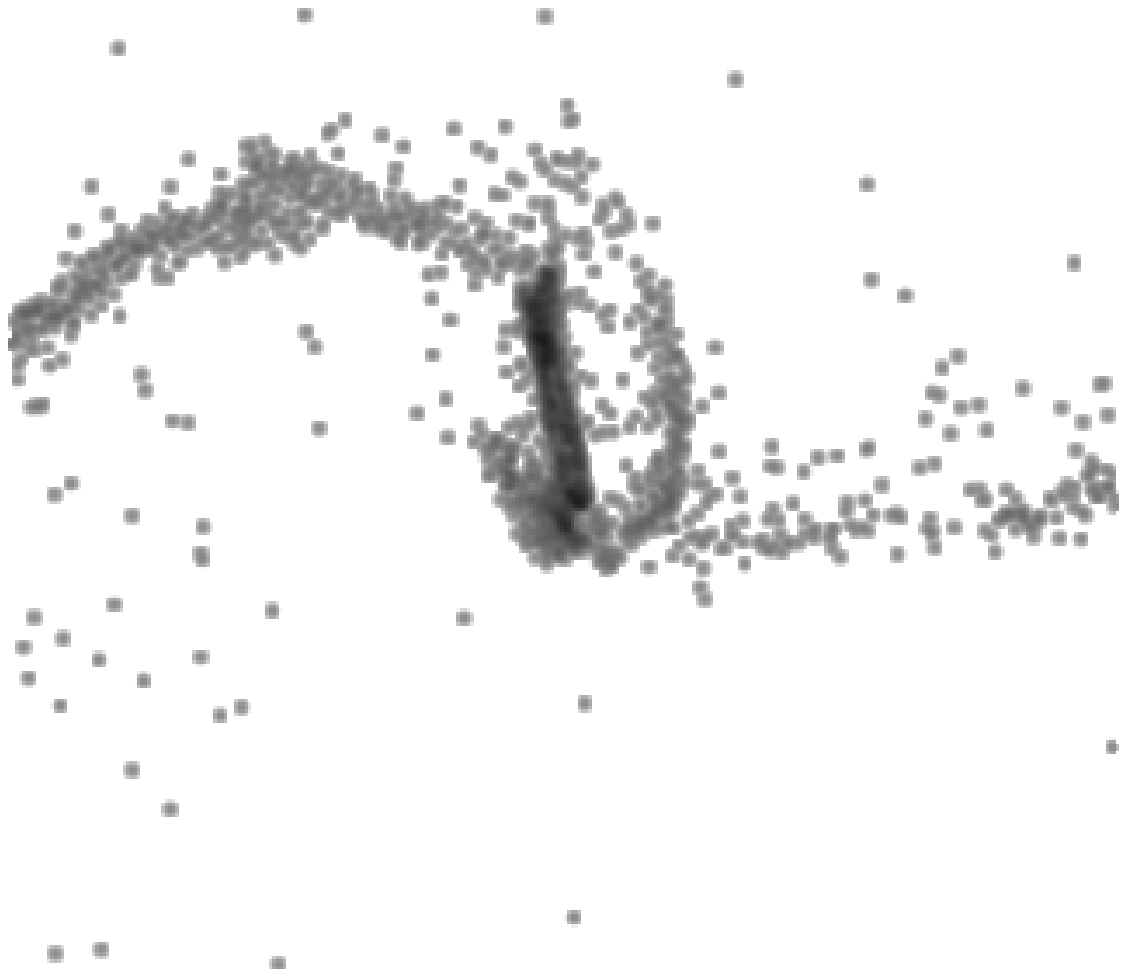, height=2.5 true cm, angle=-90}
\epsfig{figure=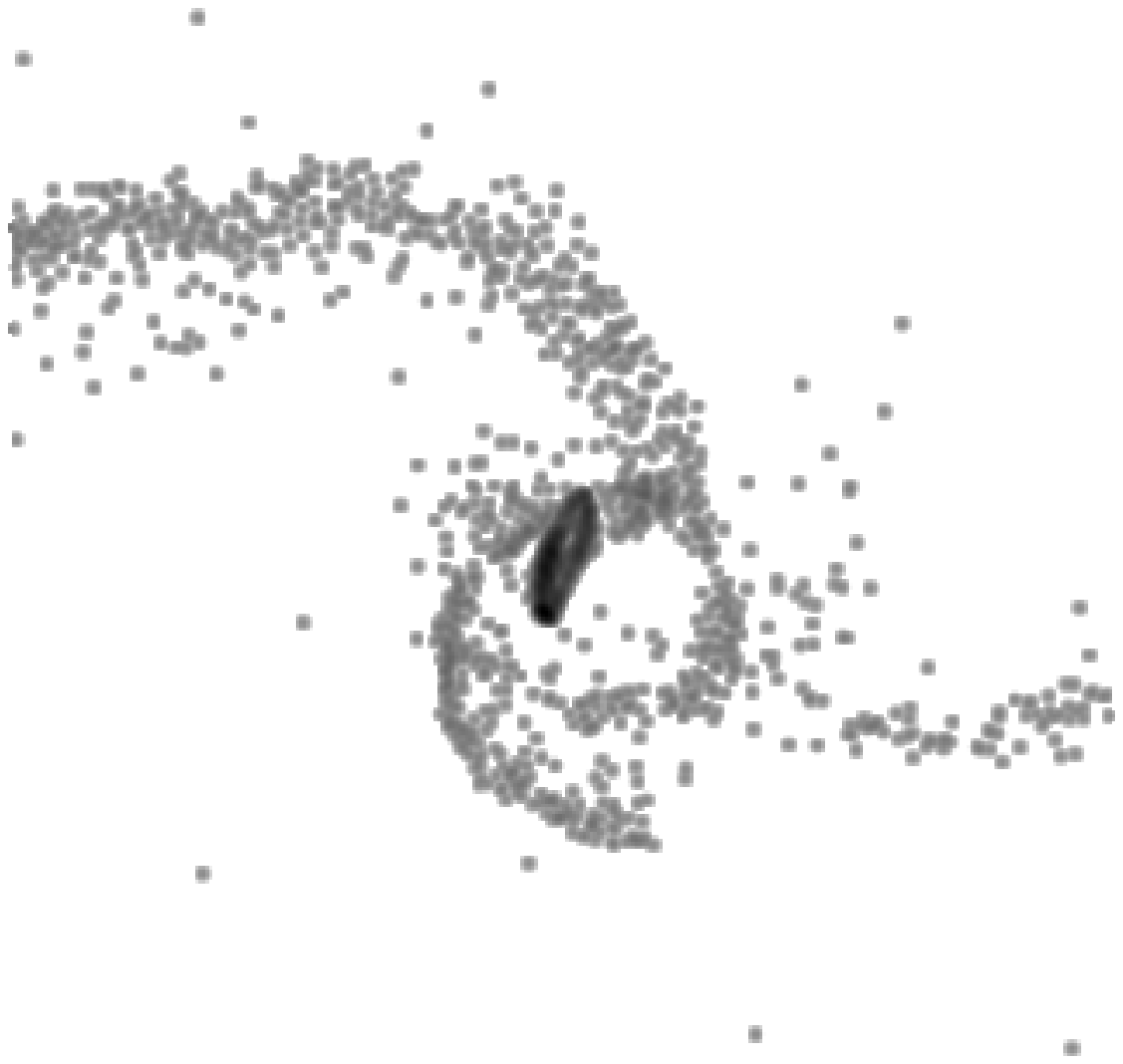, height=2.5 true cm, angle=-90}
\epsfig{figure=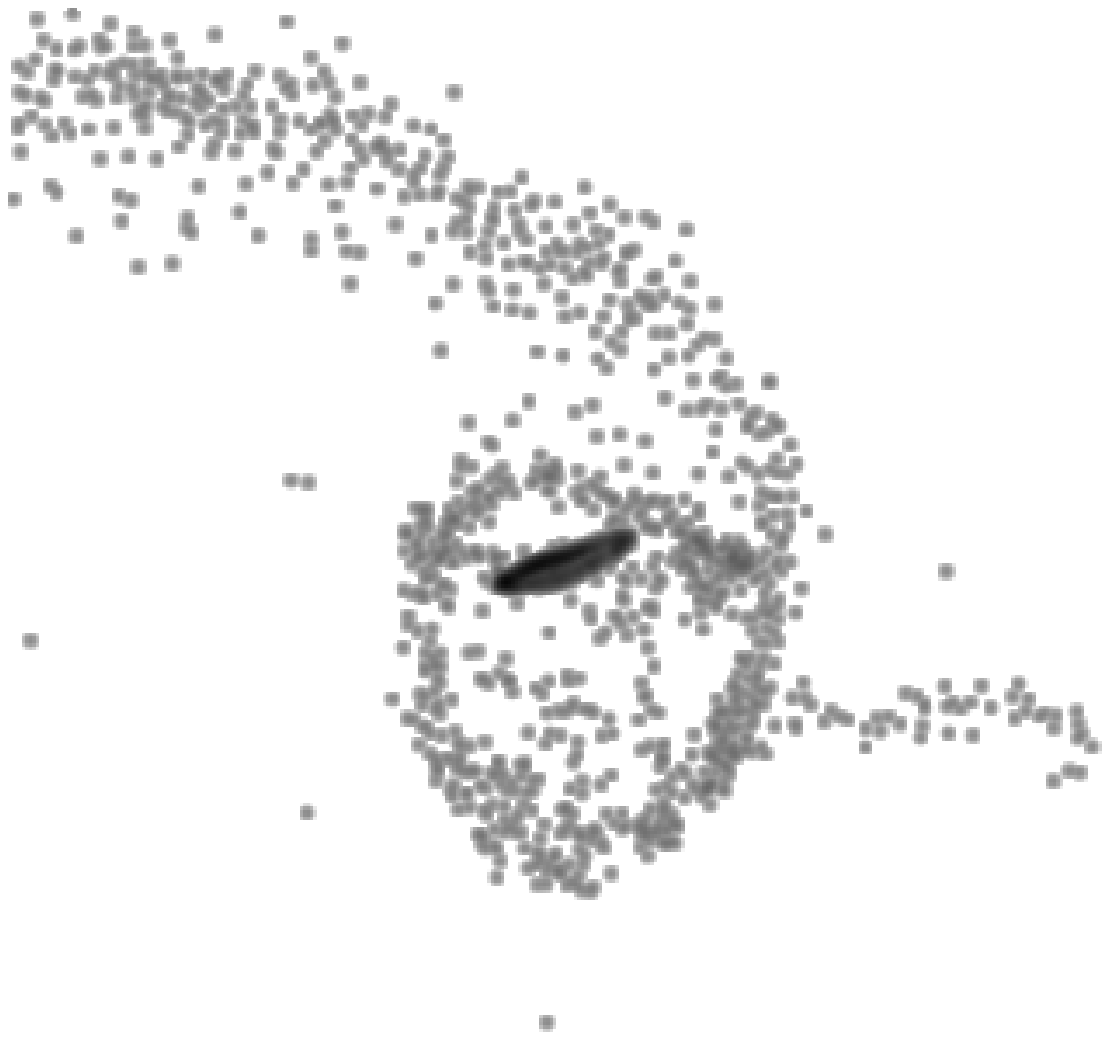, height=2.5 true cm, angle=-90}
\epsfig{figure=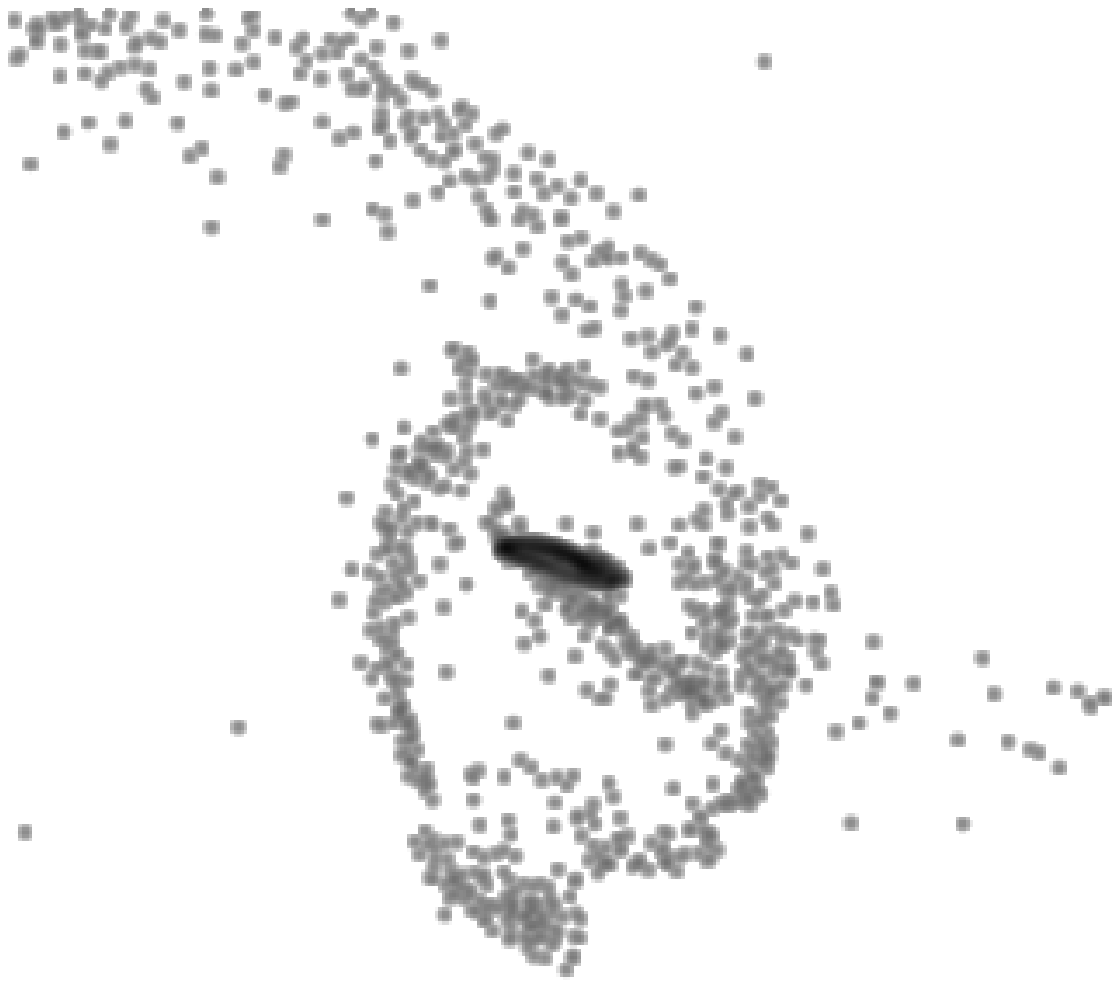, height=2.5 true cm, angle=-90}
\epsfig{figure=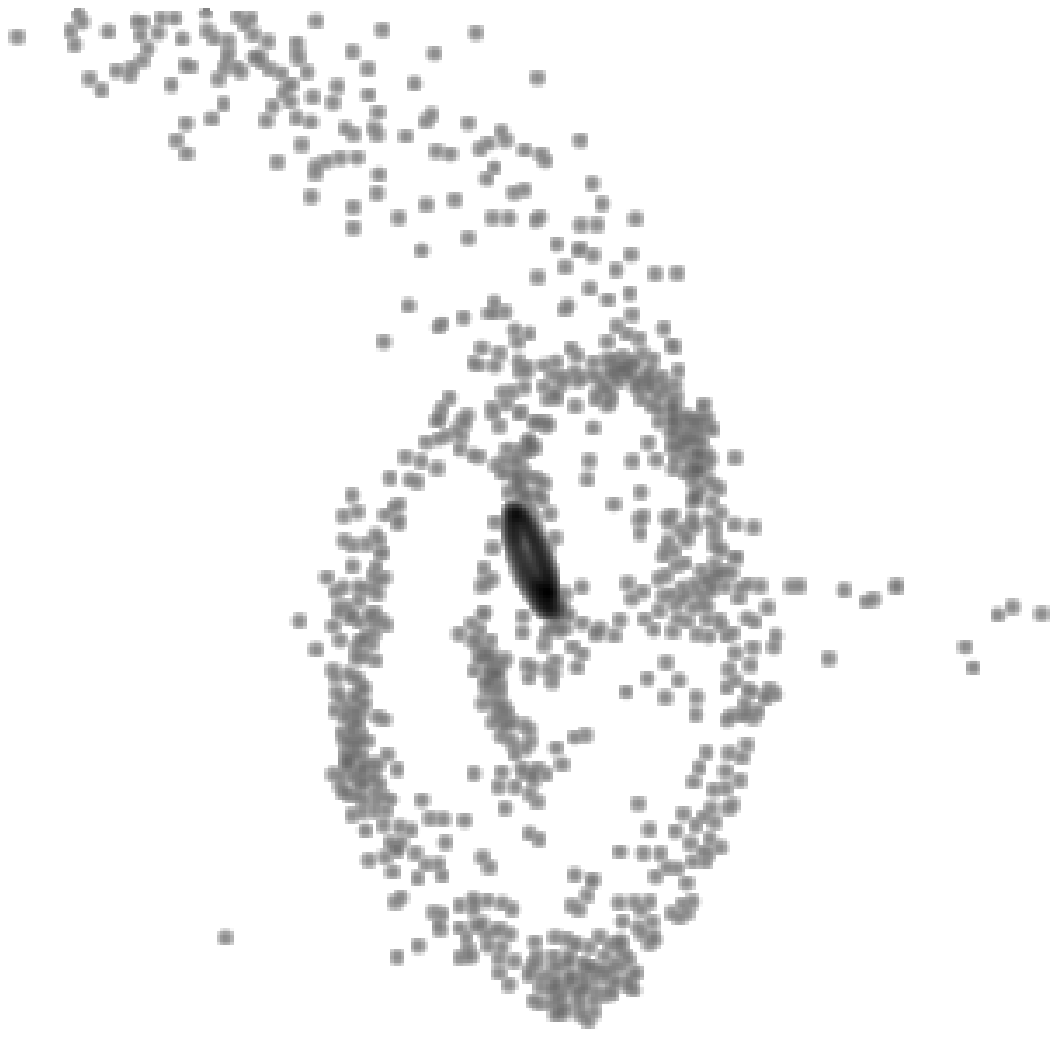, height=2.5 true cm, angle=-90}
\end{center}
\caption{Formation of a gas ring by accretion of material from a tidal
tail (see Barnes \& Hernquist 1998, video segment 5, section 2).}
\label{ringform}
\end{figure}

Goals three and four depend on gas dynamics.  In high-resolution HI
maps, gas in the southern tail seems to join continuously onto the
ring in NGC~4038.  Rings of similar size and morphology may arise as a
result of gas falling back along tidal tails; the formation of such a
ring is illustrated in Figure~\ref{ringform}.  Simulations of the
Antennae reproducing this feature might shed some light on the
conditions of star-formation in this system.  Perhaps more challenging
is to account for the IR-luminous overlap region.  This seems to be
more than just the superposition of two disks; it is probably some
sort of bridge, perhaps extended along the line of sight.

\subsection{Halo mass and orbit decay}

Detailed simulations of systems like the Antennae may also provide
useful constraints on dark halos.  To produce proper tidal tails, disk
material must escape to infinity.  Very massive, quasi-isothermal
halos prevent interacting galaxies from forming long tails (Dubinski,
Mihos, \& Hernquist 1996); clearly, such halos are not present around
galaxies like NGC~4038/9 or NGC~7252 (Mihos, Dubinski, \& Hernquist
1998).  But equally massive halos with density profiles falling off as
$\rho \propto r^{-3}$ at large $r$ are {\it not\/} excluded, as N-body
experiments explicitly demonstrate (Springel \& White 1998, Barnes
1999).  In sum, tail length tells us something about the structure of
halos, but little about their total mass.

However, it seems unlikely that arbitrary amounts of dark mass can be
included in simulations of interacting systems.  The orbital evolution
of a pair of galaxies is largely governed by the interaction of their
dark halos (White 1978, Barnes 1988).  Too much or too little orbital
decay will hinder the construction of models which evolve from
plausible initial conditions to configurations matching the observed
morphologies and velocity fields of real systems.  Possible indicators
of halo mass in interacting systems include:

\begin{enumerate}

\item Tail kinematics; the run of velocities along a tidal tail may
constrain the potential.

\item Tail fallback; if orbit decay is strong, returning tail material
may miss the disk by a wide margin.

\item Galaxy velocities; do the hulks preserve their original sense of
motion about each other?

\end{enumerate}

\noindent
The last of these, in particular, seems relevant to NGC~4038/9; the
galaxies must retain a good deal of their orbital angular momentum to
produce the crossed tails emblematic of this system.

\section{Dissipation and Thermodynamics}

Unlike stars, gas responds to pressure forces as well as gravity;
moreover, gas flows develop shocks whereas streams of stars freely
interpenetrate.  Even without the complications of star formation, the
dynamics of gas in interacting galaxies is a difficult problem.  But
dissipative dynamical systems generally possess {\it attractors\/}; in
the long run, most trajectories are captured by one attractor or
another.  Consequently, gas in interacting galaxies tends to end up in
a few stereotypical structures.

The thermodynamic history of the gas is probably the factor which
determines its fate.  To date, most simulations treat gas
thermodynamics rather crudely; the cooling function is cut off at
$T_{\rm c} = 10^4 {\rm\,K}$ to prevent the gas from ``curdling'', and
the resulting behavior is basically that of an isothermal fluid with
$T = T_{\rm c}$ (Barnes \& Hernquist 1996, hereafter BH96).  Improving
on the present treatments may require including star formation and
feedback; one possible approach to this difficult problem is described
in \S~4.  The rest of this section reviews results obtained with and
without cooling in an attempt to anticipate the results of more
realistic experiments.

Work by several investigators confirms that tidal perturbations of
gas-rich disk galaxies result in rapid gas inflows (Icke 1985, Noguchi
1988, Hernquist 1989, Barnes \& Hernquist 1991).  The immediate
physical cause of these rapid inflows is a systematic transfer of
angular momentum from the gas to the disk stars; tidally perturbed
disks develop bars (or other non-axisymmetric structures) which exert
gravitational torques on the gas (Combes, Dupraz, \& Gerin 1990,
Barnes \& Hernquist 1991).  Such inflows require strong shocks and
rapid cooling, which work together to drive the gas irreversibly
toward the center of the potential.  As a perturbed disk settles down
the gas often converges onto kpc-scale closed orbits aligned with the
stellar bar (BH96).

Dissipative mergers between disk galaxies lead to further inflows,
with large amounts of gas collecting in $\sim 0.2 {\rm\,kpc}$-scale
clouds (Negroponte \& White 1983, Barnes \& Hernquist 1991).  These
nuclear clouds contain material driven in toward the centers of
galaxies by earlier tidal interactions; during the final merger, the
gas again loses angular momentum to the surrounding material.  It
seems likely that the same physical mechanism lies behind the inflows
in perturbed disks and in merger remnants; in both cases the entropy
of the system grows as gas falls inwards.

A different fate awaits the gas which does not suffer strong shocks
and subsequent cooling in the early stages of an encounter.  This
material does not participate in rapid inflows, and retains much of
its initial angular momentum.  Consequently, it tends to collect in an
extended, rotationally supported rings or disks; one such example has
already been presented in Figure~\ref{ringform}.  In merger remnants,
such disks may be strongly warped by gas falling back from tidal tails
(BH96).  Early-type galaxies with warped gas disks include NGC~4753
(Steiman-Cameron, Kormendy, \& Durisen 1992) and NGC~5128 (van Gorkom
et al.~1990); though these disks are usually attributed to accretions
of gas-rich satellite galaxies, some may actually result from major
mergers.

The two outcomes just described -- nuclear clouds or extended disks --
seem to be the only real attractors available to dissipative gas in
merger simulations.  However, if the gas fails to cool then another
outcome is likely -- a pressure-supported atmosphere about as extended
as the stellar distribution (BH96).  Though most phases of the ISM
cool efficiently, initially hot gas ($T \ga 10^5 {\rm\,K}$, $n \la
10^{-3} {\rm\,cm^{-3}}$) could be shock-heated during a merger and
might produce envelopes of X-ray gas like those found around some
elliptical galaxies.  On the other hand, X-ray observations of the
Antennae (Read, Ponman, \& Wolstencroft 1995) and Arp~220 (Heckman et
al.~1996) reveal apparent {\it outflows\/} of up to $10^9
{\rm\,M_\odot}$ of hot gas.  The properties of these outflows are
inconsistent with shock-heating and seem to require significant
injections of mass and energy from merger-induced starbursts.

\section{Structure of Merger Remnants}

Much of this meeting has focused on possible ways in which nuclear
mass concentrations -- such as steep central cusps or black holes --
influence the global structure of elliptical galaxies.  This
discussion is motivated by the apparent dichotomy (Kormendy, these
proceedings) between faint ellipticals (which have steep central
profiles, ``disky'' isophotes, and rapid rotation) and bright
ellipticals (which have shallow central profiles, ``boxy'' isophotes,
and slow rotation).  But other factors besides central density profile
can influence galaxy structure.

\subsection{Unequal-mass mergers}

Rapidly-rotating systems may result when a large disk galaxy merges
with a smaller companion, as illustrated in a modest survey of
unequal-mass encounters (Barnes 1998).  In these experiments, both
galaxies contained bulges, disks, and halos; the larger galaxy had $3$
times the mass of the smaller, and rotated $\sim 1.32$ times faster.
The galaxies were launched on initially parabolic orbits and went
through several passages before merging; remnants were evolved for
several more dynamical times before being analyzed.

\begin{figure}[b!]
\begin{center}
\epsfig{figure=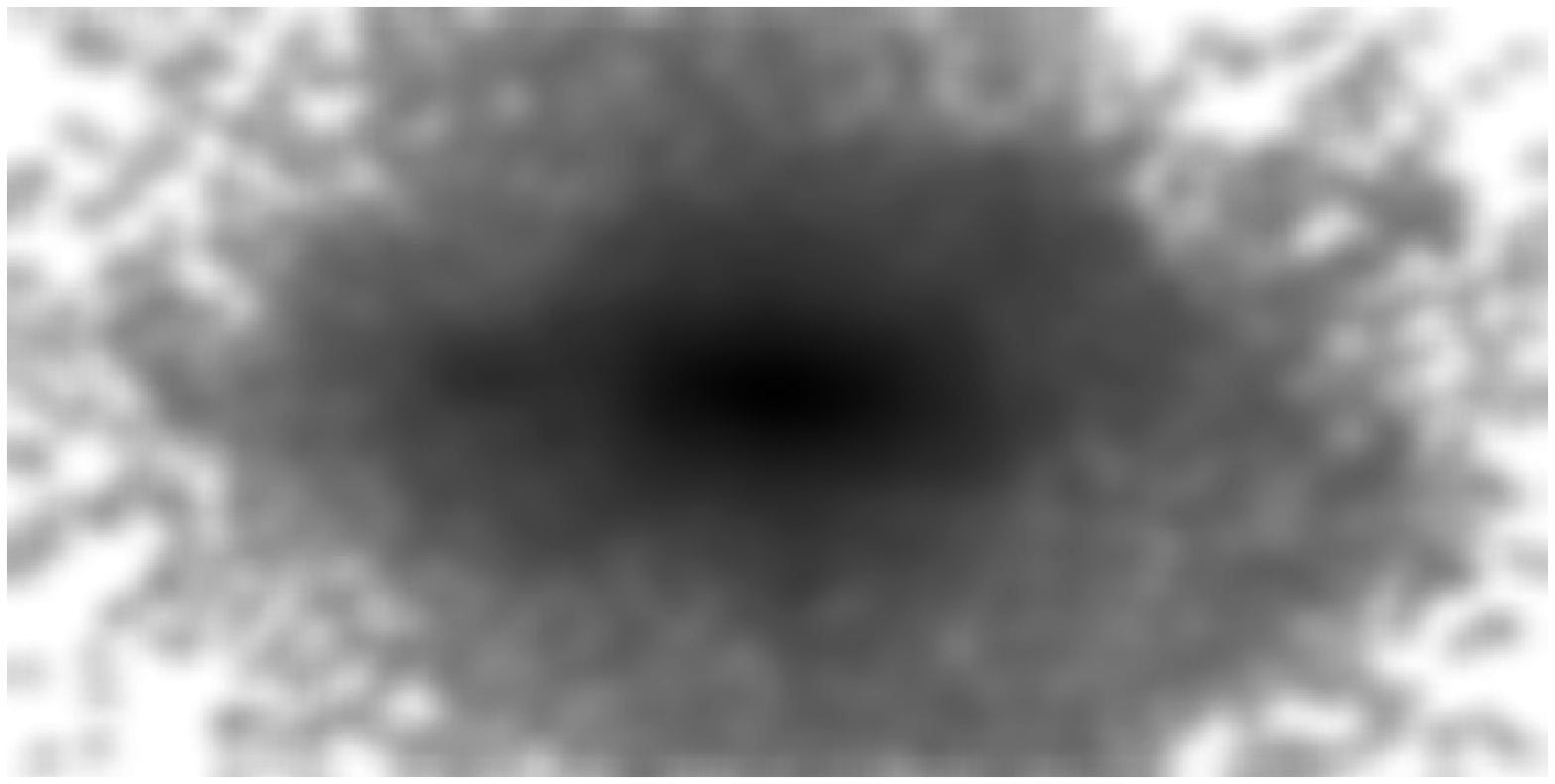, height=3.0 true cm}
\epsfig{figure=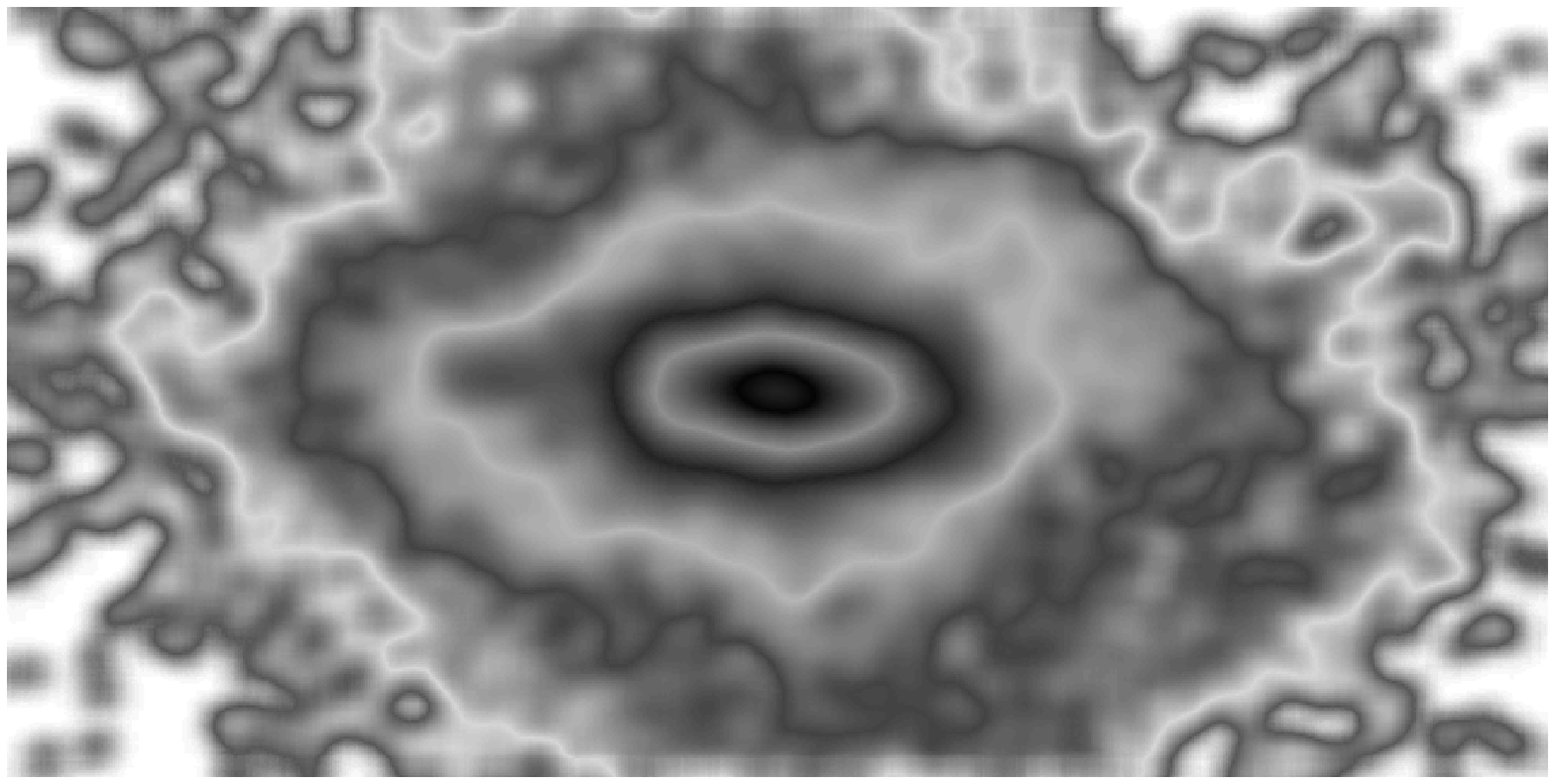, height=3.0 true cm}
\end{center}
\begin{center}
\epsfig{figure=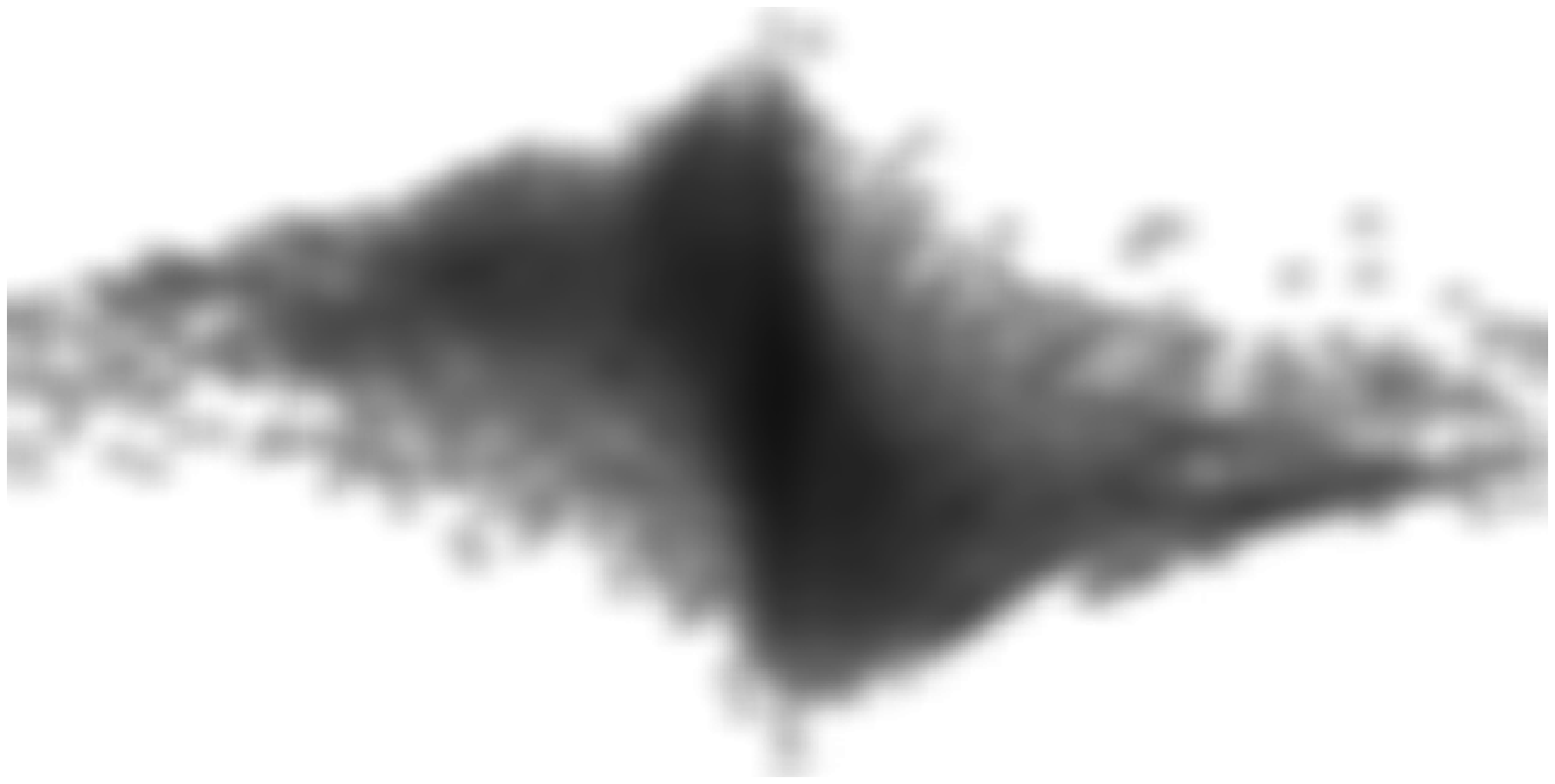, height=3.0 true cm}
\epsfig{figure=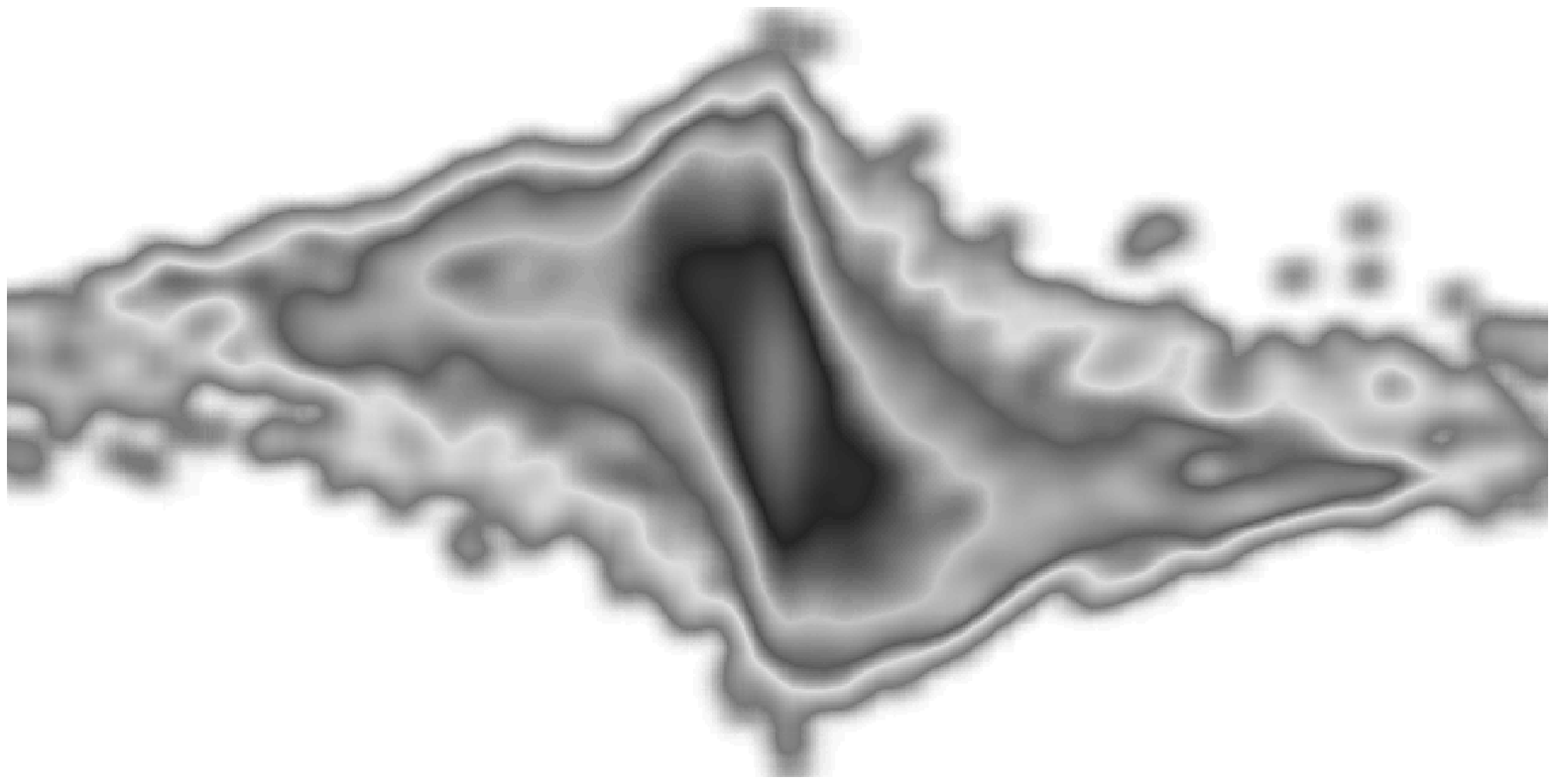, height=3.0 true cm}
\end{center}
\caption{Remnant produced by an unequal-mass merger.  Top: edge-on
views.  Bottom: line-of-sight velocities versus major axis position.
Left: natural grey scale.  Right: cyclic grey scale.}
\label{remprof}
\end{figure}

Figure~\ref{remprof} shows edge-on views and velocity distributions
for an unequal-mass merger remnant.  Unlike the products of equal-mass
mergers (Barnes 1992), this object is fairly oblate, with axial ratios
$b/a \simeq 0.9$, $c/a \simeq 0.6$.  A good deal of ``fine structure''
is still present due to incomplete phase-mixing, but the edge-on views
show a distinctly disky morphology.  The velocity profiles, which
mimic the result of placing a narrow slit along the remnant's major
axis, show that this object is rapidly rotating; in fact, $v/\sigma
\ga 2$ or more at larger radii.  And as the cyclic version of the
velocity plot makes clear, the line profiles are asymmetric, rising
sharply on the leading side of the peak, but falling off gradually on
the trailing side.

The initial conditions used for this experiment were fairly generic;
the larger disk was inclined by $i = 71^\circ$ with respect to the
orbital plane, the smaller disk by $i = 109^\circ$.  Between half and
three-quarters of a small sample of unequal-mass merger remnants
exhibit the rapid rotation and asymmetric line profiles seen in this
example (Bendo \& Barnes, in preparation).  These objects clearly
don't resemble bright ellipticals, but do seem fairly similar to faint
ellipticals or S0 galaxies.  The morphological and kinematic features
which support this classification are due to the incomplete scrambling
of galactic {\it disks\/} in unequal-mass mergers.  If dissipative
collapse is responsible for the rapid rotation of faint ellipticals
(Kormendy 1989), a subsequent merger may still be needed to {\it
transform\/} the resulting object into something resembling an
early-type galaxy.

\subsection{Dissipative mergers}

By producing inflows, dissipation can dramatically deepen galactic
potential wells, and these deeper wells seem to influence the dynamics
of collisionless material (eg., Katz \& Gunn 1991, Udry 1993, Dubinski
1994, BH96).  But these studies mostly examined effects of dissipation
on dark halos; only the last one focused on disk-galaxy mergers, and
that work compared but one pair of carefully-matched simulations.

The two remnants compared by BH96 were produced by mergers of
equal-mass bulge/disk/halo galaxies.  Both experiments started with
{\it exactly\/} the same initial conditions, using disk inclinations
of $0^\circ$ and $71^\circ$; both were evolved with the same spatial
resolution (a.k.a.~``force softening'').  In the dissipative version,
a tenth of the disk mass was treated as gas with a cooling cut-off at
$T_{\rm c} = 10^4 {\rm\,K}$, while in the collisionless version
everything obeyed the collisionless Boltzmann equation.

\begin{figure}[b!]
\begin{center}
\epsfig{figure=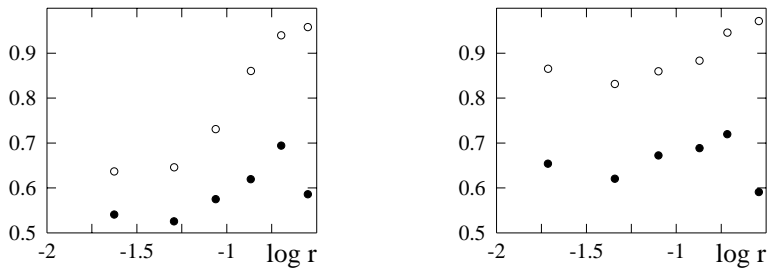, height=3.5 true cm}
\end{center}
\caption{Ellipticity profiles for collisionless (left) and dissipative
(right) versions of the same merger remnant.  Open circles represent
$b/a$, filled circles $c/a$.}
\label{elliprof}
\end{figure}

Figure~\ref{elliprof} compares the ellipticity profiles of these two
remnants.  Beyond their half-light radii ($r_{\rm hl} \simeq 0.18$
model units) both remnants are nearly oblate and rotate rapidly in
memory of the direct ($i = 0^\circ$) disks used in the initial
conditions.  But inside $r_{\rm hl}$ the two remnants are quite
different; the collisionless version is a triaxial ellipsoid rapidly
tumbling about its minor axis, while the dissipative version is fairly
oblate and slowly rotating.

How does dissipation influence the shape of merger remnants?  The
dissipative remnant has a deeper potential well as a result of its
central gas cloud, which contains $\sim 4.5$\% of the luminous mass,
or $\sim 0.9$\% of the total.  But the finite resolution of the force
calculation spreads this central mass over a radius of $\sim 0.04 \,
r_{\rm hl}$; thus compared to a black hole or singular logarithmic
potential, this mass may be ineffective at scattering box orbits
(Valluri, these proceedings).  Moreover, the oblate shape of the
remnant seems to be established at the moment of the merger itself
instead of developing progressively from the inside out (Ryden, these
proceedings).

Thinking that the shapes of these remnants might be constrained by the
scarcity of box orbits, I constructed a composite mass model with the
density profile of the dissipational remnant and the ellipticity
profile of its collisionless counterpart, and used its potential to
evaluate the phase-space volumes of the major orbit families (Barnes
1998).  While this composite offered fewer boxes and more z-tubes
than the collisionless remnant, bona-fide box orbits were present at
all binding energies.  Thus self-consistent equilibria as centrally
concentrated as the dissipational remnant and as flattened as the
collisionless remnant may exist.  However, some finesse is probably
required to realize such equilibria.  Merging sows stars far and wide
across phase space; not all physically consistent systems may be
constructed with such a blunt instrument.

All of this work is based on only one pair of simulations, and the two
remnants compared by BH96 may not be entirely typical.  For example,
the pre-merger disks in these experiments developed bars, and the bars
in the dissipational version had significantly higher pattern speeds.
Thus when the disks merged, their bars had different orientations, and
this might influence remnant structure.  Comparison of a larger sample
of collisionless and dissipative merger remnants is clearly warranted,
but sufficient computer power is hard to find.  Meanwhile,
collisionless mergers between models of various central concentrations
may help expose the connection between density profile and remnant
shape (Fulton \& Barnes, in preparation).

\section{Simulations of Starburst Galaxies}

The crude treatment of gas thermodynamics in most work to date is
perhaps the greatest barrier to simulating star formation in
interacting galaxies.  As described in \S~2, radiative cooling is
typically cut off at $10^4 {\rm\,K}$, and most of the gas remains
close to this temperature.  Stars, on the other hand, form at much
lower temperatures; consequently, sites of star formation can't be
directly located in the simulated gas.

Within the framework of most simulations, gas density is the only
variable with an interesting range of values, so most treatments
assume the star formation rate is a function of the gas density.  This
approach has some justification; studies of star formation in systems
ranging from quiescent disk galaxies to violent starbursts find that
star formation rates roughly follow a Schmidt (1959) law of the form
$\dot\Sigma_s \propto \Sigma_g^n$, where $\Sigma_s$ and $\Sigma_g$ are
the stellar and gaseous surface densities, respectively, and the index
$n \simeq 1.4 \pm 0.15$ (eg., Kennicutt 1998).  The usual approach is
thus to adopt a star formation law of the form $\dot\rho_s \propto
\rho_g^n$, where $\rho_s$ and $\rho_g$ are the stellar and gaseous
volume densities, respectively.

The implementation of feedback effects due to stellar evolution and
supernovae is particularly difficult.  Cooling is so rapid that the
otherwise plausible strategy of dumping thermal energy into the gas
proves ineffective; the energy is radiated away before anything else
can happen (Katz 1992, Summers 1993).  Another trick is to impart some
outward momentum to gas particles surrounding sites of star formation
and/or supernovae; this seems more effective, but involves an
arbitrary efficiency factor (Navarro \& White 1993, Mihos \& Hernquist
1994).  It's unlikely that feedback can be properly implemented as
long as the gas is effectively treated as a single-phase medium.

A promising alternative to density-driven star formation is now
available (Gerritsen \& Icke 1997).  In this approach the gas is
allowed to cool below $10^4 {\rm\,K}$, and sites of star formation are
defined by a Jeans criterion.  The stellar radiation field, calculated
in the optically thin limit, is used to heat the gas.  Star formation
is thus a self-regulating process; negative feedback maintains the
system in a quasi-stable state while slowly converting gas to stars.
Competition between radiative heating and cooling creates a two-phase
medium with temperatures of $10^2 {\rm\,K}$ and $10^4 {\rm\,K}$; a
third phase at $10^6 {\rm\,K}$ appears when the effects of supernovae
are included.  As a bonus, the resulting star formation obeys a
Schmidt law with index $n \simeq 1.3$.

It may turn out that many of the desirable features of this approach
are simple consequences of combining radiative cooling and negative
feedback.  Some details surely require modification; the treatment of
the radiation field seems particularly suspect since galactic disks,
edge-on, are not optically thin.  But the general view of star
formation as a self-regulating process and the re-introduction of gas
temperature as a physically interesting variable are surely major
improvements on previous treatments.

Does the treatment of star formation make a real difference in the
outcome of simulations?  In at least one respect, it does.
Simulations using the Schmidt law predict that interacting late-type
disk galaxies consume most of their gas shortly after their first
passage; merger-induced starbursts only result if the disks are
protected from bar formation by compact central bulges (Mihos \&
Hernquist 1996).  In contrast, simulations using self-regulated star
formation predict that bulgeless disk galaxies retain enough gas to
fuel ultra-luminous starbursts during their final mergers; while star
formation rates increase after the first passage, radiative heating
delays violent star formation until the merger drives most of the gas
into a compact central cloud (Gerritsen 1997).

To date, outflows like those seen in interacting starburst galaxies
have not been reproduced with either treatment of merger-induced star
formation.  This remains an important challenge for the future.

\section{Discussion}

Within the space of this brief review, it's impossible to do more than
touch on a few aspects of galaxy interactions; I've said nothing about
the tidal genesis of grand-design spirals, origins of ring galaxies,
multiple mergers and the formation of cD galaxies, or interactions in
groups and clusters, to name a few.  The main points I have tried to
address are listed here:

1. Simulating interacting galaxies is an {\it art\/}; it can't be
reduced to a recipe.  Picasso defined art as ``a lie which makes us
realize truth'', and this seems to be a good stance to adopt when
trying to reproduce real galaxies.  Some features of the Antennae may
be clues leading to fundamental insights, while others may be due to
quirks of the pre-encounter disks.  We don't always know which is
which; experience is the only guide.

2. In interacting galaxies, thermodynamics is the key to the fate of
the gas.  Gas which encounters strong radiative shocks in the early
phases of a collision will diverge from the stars and fall into the
centers of interacting galaxies and merger remnants.  Gas which does
not suffer such shocks until the later stages of a collision, on the
other hand, retains much of its initial angular momentum and builds up
extended disks.

3.  Remnant structure is determined by many factors.  Steep central
cusps (or nuclear black holes) may suppress strong triaxiality, but
this doesn't explain why galaxies with such profiles rotate rapidly.
More generally, sheer {\it existence\/} of self-consistent equilibria
is not enough to explain the properties of elliptical galaxies; the
details of formation play an important role.

4. Simulations including star formation are still in their early days;
a good deal of further work is needed to develop and test alternate
approaches.  Plausible treatments of feedback from star formation and
evolution require abandoning the assumptions which effectively limit
the simulated gas to a single phase.

As noted in the abstract, galaxy mergers have deep connections to
galaxy formation.  For example, the issues reviewed in \S~2 and~4
arise in cosmological simulations of disk galaxy formation; in
dissipative CDM simulations, gas inflows are so efficient that little
remains to build disks (Navarro \& Benz 1991, Navarro \& White 1994,
Navarro \& Steinmetz 1997).  The resolution of this problem is
probably to implement strong feedback of the kind long assumed in
hierarchical models of galaxy formation (White \& Rees 1978, White \&
Frenk 1991, Kauffmann, White, \& Guiderdoni 1993, Navarro, Frenk, \&
White 1995).  This may well be the same sort of feedback needed to
reproduce the outflows of hot gas in violently interacting starburst
galaxies.

\acknowledgements

I thank John Hibbard for allowing me to discuss our unpublished work
on NGC~4038/9 and Lars Hernquist for providing me with a copy of
TREESPH.  I'm also grateful to Jun Makino and the University of Tokyo
for hospitality while I was writing this report.  This research has
made use of NASA's Astrophysics Data System Abstract Service.

\end{document}